 \documentclass[12pt]{article}
\usepackage{graphicx}

\begin{document}

%\draft
\title{
Effect of symmetry of the electron states of HTSC on the
current-voltage characteristics of SIS junctions. }
\author{P.I.Arseev., N.K. Fedorov, S.O. Loiko\\
%\address{
P.N. Lebedev Physical Institute, Russian Academy of Science.\\
Leninskii pr. 53, Moscow, 119991 Russia.}
\date{}
\maketitle

\begin{abstract}
The current-voltage $IV$ characteristics of $SIS$ junctions are
calculated in the framework of a multiband model with an
anisotropic effective order parameter of HTSC. The results of
calculations show that the shape of $IV$ characteristic and the
density of electron states changes significantly depending on the
parameters of the model. A theoretical explanation is proposed for
the experimentally observed $s$-like behavior of the $IV$
characteristics of $SIN$ and $SIS$ junctions with BSCCO-type
superconductors. The dependence of the superconducting peaks
asymmetry on the mutual arrangement of the bands is analyzed. The
difference between the obtained results and the results of
single-band models with the $s$ and $d$ symmetries of the order
parameter is discussed.
\end{abstract}

The lack of a generally accepted explanation of the pairing
mechanism in high-temperature superconductors (HTSC) makes the
interpretation of numerous experimental data (including the
results of tunnel experiments) difficult. The complexity of the
crystal structure of HTSC compounds also complicates the
situation. In most cases, it is difficult to explain peculiarities
in the tunnel characteristics within the framework of the standard
Bardeen-Cooper-Schrieffer (BCS) model. Such peculiarities include
the variety of the subgap structure ranging from a linear $d$-type
structure to a nearly planar structure of the $s$ type
\cite{zasad}, the asymmetry of superconducting peaks at the edge
of the gap \cite{kuz,ren}, and nontrivial behavior of the tunnel
density of states of $SIN$ junctions and of the current-voltage
characteristics of $SIS$ junctions outside the gap region. In some
experiments \cite{pon,suz}, strongly suppressed
$\displaystyle\frac{dI}{dV}$ characteristics with very narrow
peaks typical of $s$-type superconductors were obtained in the
subgap region, which contradict at first glance the available data
on strong anisotropy of the order parameter in the $CuO_2$ plane
\cite{ding}.

A large number of theoretical models have been proposed to explain
the variety of experimental data. These include the models based
on various features of the band structure (taking into account,
for example, the closeness of the Van Hove singularities to the
chemical potential \cite{wei,fed}), on the choice of the order
parameter symmetry \cite{fed}, and on the properties of the tunnel
matrix element \cite{led} in the single-band BCS scheme. Even the
slave-boson approach \cite{slave} was used for explanation the
asymmetry of the superconducting peaks and the inhomogeneity of
the density of electron states in the normal state. A series of
publications \cite{pines,bulut,cof,schl}
 is devoted to the role of inelastic scattering, including
the inelastic scattering by antiferromagnetic spin fluctuations.
The results of theoretical investigations show that the inclusion
of Van Hove singularities and the use of the $(s + d)$ symmetry of
the order parameter are obviously essential for a correct
explanation of the most experimentally observed properties of the
tunnel conductivity of HTSC. For this reason, the experimental
investigation of the properties of the electron spectrum of
high-temperature superconductors in the normal state is very
important.

In our opinion, it is the analysis of the crystal structure that
makes it possible to explain a number of results of tunnel
measurements in high-temperature superconductors. This paper aims
at explanation the features of tunnel characteristics using the
model with an anisotropic effective order parameter, which is
based only on the properties of the electron spectrum of HTSC
\cite{ars}. The spectrum of the considered model approximately
corresponds to the band structure of compounds of the BiSrCaCuO
(BSCCO) type. An important feature of this approach is that the
initial electron-electron interaction leading to pairing is
treated as isotropic in the $CuO_2$ plane. The anisotropy of the
order parameter and of the excitation spectrum is determined by
the symmetry properties of the crystal lattice. According to our
calculations, many features of the tunnel characteristics typical
of HTSC compounds
 can be explained by using a rather universal approach which
will be described below.

Let us consider the effect of symmetry of the initial bands of the
superconductor on the current-voltage characteristics of
$SIS$-junctions. The electron system of the $CuO_2$ plane is
described by model Hamiltonian of the form:

\begin{eqnarray}
H & \quad = \quad & \varepsilon_{z^2}\sum_{i,\sigma }c_{i,\sigma
}^{+}c_{i,\sigma }
       +\varepsilon_{d} \sum_{i,\sigma }
d_{i,\sigma }^{+}d_{j,\sigma } + \varepsilon_{p}\sum_{i,\sigma
}(p_x^{+}(i,\sigma )p_x(i,\sigma ) + p_y^{+}(i,\sigma
)p_y(i,\sigma )) +
                                                 \nonumber \\
       & \quad + \quad & \sum_{i,j,\sigma }
  (t_{z^2-p_x}^{i,j} c_{i,\sigma }^{+}p_x(j,\sigma ) + h.c.)
+ \sum_{i,j,\sigma } (t_{z^2-p_y}^{i,j} c_{i,\sigma }^{+}p_y(j,\sigma ) + h.c.)
 +
                                           \nonumber \\
& \quad +   \quad &
\sum_{i,j,\sigma }  (t_{d-p_x}^{i,j} d_{i,\sigma }^{+}p_x(j,\sigma ) + h.c.)
 +
  \sum_{i,j,\sigma } (t_{d-p_y}^{i,j}d_{i,\sigma }^{+}p_y(j,\sigma ) + h.c.)
                                         +       \nonumber \\
 & \quad + \quad &
    U_{z^2}\sum_{i}c_{i\downarrow}^{+}c_{i\downarrow}
 c_{i\uparrow}^{+}c_{i\uparrow}
+U_d\sum_{i}d_{i\downarrow}^{+}d_{i\downarrow}d_{i\uparrow}^{+}
      d_{i\uparrow}  \qquad ,
%                               \eqnum{1}
\end{eqnarray}
 Here
$c^{+}_{i,\sigma }$
  and
$d_{i,\sigma }^{+}$ are creation operators for electrons with spin
 $\sigma$ on the $d_{z^2}$- and
$d_{x^2-y^2}$- orbitals of the $i$-th site of copper;
$p_x^{+}(i,\sigma )$
  and
$p_y^{+}(i,\sigma )$ are the creation operators for electrons with
spin $\sigma$ on the $p_x$ and $p_y$ orbitals of the $i$-th site
of oxygen (Fig.1). $\varepsilon_p$, $\varepsilon_d$ and
$\varepsilon_{z^2}$ are the energies of the $p$ levels of oxygen
and of the  $d_{x^2-y^2}$ and $d_{z^2}$ levels of copper,
respectively, measured from the chemical potential ($\mu=0$).
$t_{z^2-p_x}^{i,j}=t_{z^2-p_y}^{i,j} \equiv t_{z^2-p}$ and
$t_{d-p_x}^{i,j}=-t_{d-p_y}^{i,j} \equiv t_{d-p}$ are the matrix
elements of one-particle transitions between the
  $d_{z^2}$ and $d_{x^2-y^2}$ orbitals of copper and the $p$ orbitals of oxygen.
For the sake of simplicity, we assume that superconductivity is
caused by the isotropic attraction $U_{z^2}<0$ of electrons on the
$d_{z^2}$ orbital. It should be noted that a more complicated
consideration of the superconducting interaction would not lead to
a qualitative difference from the case investigated here. We also
assume the presence of the isotropic effective electron-electron
interaction $U_d$ on the $d_{x^2-y^2}$ orbital of copper.

\begin{figure}
%\centering{\epsfbox{fig1.eps}
 \centering{\includegraphics{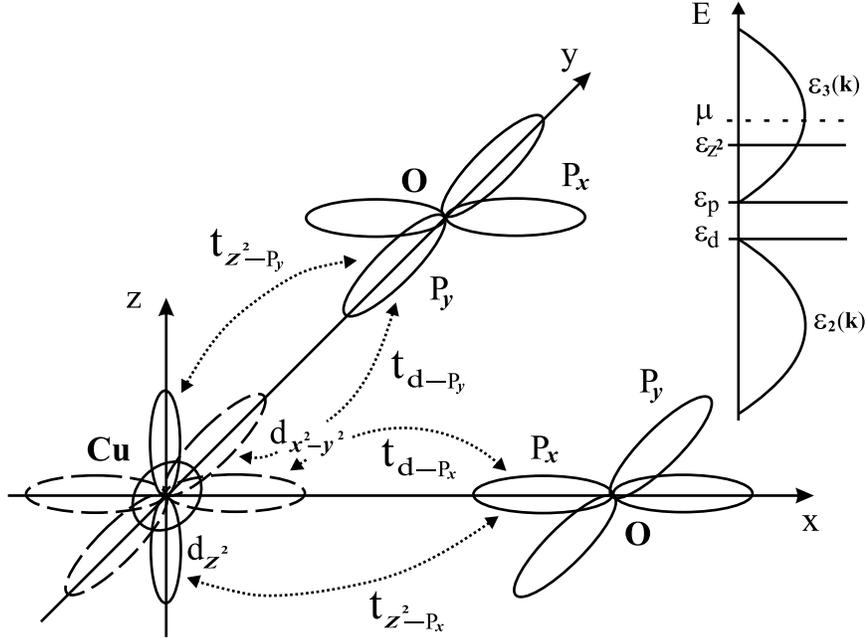}
\caption{ Schematic representation of atomic orbitals in the
$CuO_2$ plane and transitions between these orbitals included in
Hamiltonian (1). The inset shows schematically the arrangement of
energy bands and levels corresponding to Hamiltonian (4).} }
\end{figure}

Introducing new operators $a^{(\beta)}_{{\bf k}, \sigma }$
($\beta$=1,2,3) in the $\bf k$ representation according to the
formulas
\begin{eqnarray} d_{{\bf k},\sigma } = C^{(1)}_d {a^{(1)}_{{\bf k},\sigma }}+
C^{(2)}_d {a^{(2)}_{{\bf k},\sigma }}+C^{(3)}_d {a^{(3)}_{{\bf
k},\sigma }} \nonumber
\end{eqnarray}
\begin{eqnarray} p_x({\bf k},\sigma) = C^{(1)}_{p_x} {a^{(1)}_{{\bf k},\sigma }}+
C^{(2)}_{p_x} {a^{(2)}_{{\bf k},
\sigma }}+C^{(3)}_{p_x} {a^{(3)}_{{\bf k},\sigma}}
\end{eqnarray}
\begin{eqnarray}
p_y({\bf k},\sigma) = C^{(1)}_{p_y} {a^{(1)}_{{\bf k},\sigma }}+
C^{(2)}_{p_y} {a^{(2)}_{{\bf k},\sigma }}+ C^{(3)}_{p_y}
{a^{(3)}_{{\bf k},\sigma }}\,, \nonumber
\end{eqnarray} we diagonalize the
Hamiltonian part describing one-particle transition between the
$d_{x^2-y^2}$ orbitals of copper and the $p$ orbitals of oxygen.
The coefficients $C_{d}$ and $C_{p_x(p_y)}$ in formulas (2) have
the form:

\begin{eqnarray}
%{ll}
 C^{(1)}_{d} = 0, \qquad \qquad \quad  \qquad& \qquad
C^{(\alpha)}_{d} =
\displaystyle \frac{\varepsilon_p -
\varepsilon_{\alpha}}{\sqrt{(\varepsilon_p -
\varepsilon_{\alpha})^2+(t_x)^2+(t_y)^2}} ,  \nonumber \\
 C^{(1)}_{p_x(p_y)} =\displaystyle \frac{i t_{y(x)}}{\sqrt{(t_x)^2+(t_y)^2}},
  &    \qquad  \qquad
C^{(\alpha)}_{p_x(p_y)} =\displaystyle \pm \frac{t_{x(y)}}
{\sqrt{(\varepsilon_p - \varepsilon_{\alpha})^2+ (t_x)^2+(t_y)^2}}
, \nonumber
\end{eqnarray}
where  $\alpha=2,3$ and
              $t_{x(y)}= t_{d-p}
 \sin{\frac{k_{x(y)}}{2}}
\,.$
 Similarly, the matrix element of a transition between the
 $d_{z^2}$ orbitals of copper and the $p$ orbitals of oxygen in the ${\bf k}$
 representation can be written as
                   ${\widetilde t_{x(y)}}=
 t_{z^2-p} \sin{\frac{k_{x(y)}}{2}}$. Since the matrix elements
 $t^{i,j}_{d-p}$ and
$t^{i,j}_{{z^2}-p}$ depend on the indices of different sites, the
quantities  $t_{x(y)}$ and ${\widetilde t_{x(y)}}$ must be
functions of ${\bf k}$. Operators ${a^{(1)}_{{\bf
k},\sigma}}^{+}$, ${a^{(2)}_{{\bf k},\sigma}}^{+}$ and
${a^{(3)}_{{\bf k},\sigma}}^{+}$ are the creation operators for
electrons with spin $\sigma$ and quasimomentum $\bf k$ in the
bands formed by hybridized $p$ orbitals of oxygen and
$d_{x^2-y^2}$ orbitals of copper with the dispersion relation
\begin{eqnarray}\varepsilon_1=\varepsilon_p\quad,\quad
\varepsilon_{2(3)}({\bf k})=\displaystyle
\frac{\varepsilon_p+\varepsilon_d}{2} \mp \frac{1}{2}
\sqrt{(\varepsilon_p-\varepsilon_d)^2+4 (t_x^2({\bf k})+t_y^2(\bf
k))}\,. \end{eqnarray}

As a result of transformation (2), the system Hamiltonian in the
mean-field approximation takes the form:
\begin{eqnarray}
H & \quad = \quad & \sum_{{\bf k},\sigma }\varepsilon_{z^2}
c_{{\bf k},\sigma }^{+}c_{{\bf k},\sigma }
       + \sum_{{\bf k},\sigma,\beta}
  \varepsilon_{\beta}{a^{(\beta)}_{{\bf k},\sigma }}^{+}
a^{(\beta)}_{{\bf k},\sigma }
               \nonumber \\
       & \quad + \quad & \sum_{{\bf k},\sigma,\beta}
  (W_{\beta}({\bf k}) c_{{\bf k},\sigma }^{+}
a^{(\beta)}_{{\bf k},\sigma } + h.c.)
                                                \nonumber \\
 & \quad - \quad &
    \sum_{\bf k}(\Delta_{z^2} c_{{-\bf k},\downarrow}^{+}c_{{\bf k},\uparrow}^{+} + h.c)
-    \sum_{{\bf k}, \beta,\gamma }(\Delta_{\beta \gamma}
{a^{(\beta)}_{{-\bf k},\downarrow}}^{+}
{a^{(\gamma)}_{{\bf k},\uparrow}}^{+} + h.c),
%                               \eqnum{4}
\end{eqnarray}
where
\begin{eqnarray} W_1({\bf k})=-\displaystyle \frac{{\widetilde t_x}t_y+{\widetilde t_y}t_x}{\sqrt{t_x^2+t_y^2}}\,,
\qquad W_{\alpha}({\bf k})=\displaystyle \frac{t_y{\widetilde
t_y}-t_x{\widetilde t_x}} {\sqrt{(\varepsilon_p -
\varepsilon_{\alpha})^2+(t_x)^2+(t_y)^2}}\,, \qquad \alpha=2,3\,
\end{eqnarray}
and new order parameters
\begin{eqnarray}
\Delta_{\beta \gamma} ({\bf k})= C^{(\beta)}_{d}({\bf k})
C^{(\gamma)}_{d}(-{\bf k}) \Delta_d\,, & \qquad \beta,
\gamma=1,2,3\, ,
\end{eqnarray}
have been introduced, in which
\begin{eqnarray}
\Delta_d=-\frac1N U_d \sum_{\bf k}
\langle d_{{\bf k},\uparrow}d_{{-\bf k},\downarrow}\rangle\,,
& \qquad \displaystyle \Delta_{z^2}
=-\frac1N U_{z^2} \sum_{\bf k}
\langle c_{{\bf k},\uparrow}c_{{-\bf k},\downarrow}\rangle\,.
\end{eqnarray}
The arrangement of the energy bands corresponding to Hamiltonian
(4) is shown in the inset to Fig.1. In the framework of a
realistic description of  BSCCO-type compounds \cite{entel}, it is
assumed that the chemical potential   $\mu$ and the $z^2$ level of
copper locate closely to the center of the upper band
$\varepsilon_3(\bf k)$. The remaining parameters of the model are
assumed to satisfy the relations
 $\varepsilon_d<\varepsilon_p$,
$\varepsilon_p-\varepsilon_d\ll t_{d-p}$, $t_{z^2-p}\ll
|\varepsilon_d|,|\varepsilon_p|$, $|\Delta_d|<\Delta_{z^2}\ll
|\varepsilon_d|,|\varepsilon_p|$.

Since only the energy range near the chemical potential (on the
order of several $\Delta_{z^2}$) plays an important role here, we
can disregard the two lower bands $\varepsilon_1$ and
 $\varepsilon_2(\bf k)$ taking
into account the relation between parameters and restrict our
analysis to the upper band $\varepsilon_3(\bf k)$ only. In
addition to one-particle transitions included in Hamiltonian (1),
we can take into account the additional hybridization of the
$d_{z^2}$ orbital with the atoms surrounding the $CuO_2$ plane,
including $BiO$ and $SrO$ complexes of the BSSCO-type compounds as
well as direct transition between the $d_{z^2}$ orbitals of
different copper atoms. In this connection, we assume that the
initial $z^2$ band is characterized by the dispersion relation
$\varepsilon_{z^2}({\bf
k})=\varepsilon_{z^2}+t_{z^2}(cos(k_x)+cos(k_y))$
 with a width much smaller than for $\varepsilon_2(\bf k)$ and
$\varepsilon_3(\bf k)$  bands. Thus, in the chosen approximation,
the initial problem can be reduced to a two-band model in which
the initial $\varepsilon_{z^2}(\bf k)$ and the $\varepsilon_3(\bf
k)$ bands with the one-particle hybridization $W_3({\bf k})$ and
the order parameters $\Delta_{z^2}$ and $\Delta_{33}(\bf k)$ are
considered.

We introduce the following time Green's functions:
\begin{eqnarray}
G_{z^2}({\bf k};t,t')=-i\left\langle T\ c_{{\bf k},\sigma }(t)c^{+}_{{\bf k},\sigma  }(t')\right\rangle,
&\quad &
g_{\sigma \sigma'}F^{+}_{z^2}({\bf k};t,t')=-i\left\langle T\
c^{+}_{-{\bf k},\sigma}(t)c^{+}_{{\bf k},\sigma'}(t')\right\rangle, \nonumber    \\
G_{{z^2},3}({\bf k};t,t')=-i\left\langle T\ c_{{\bf k},\sigma}(t)a^{(3)+}_{{\bf k},\sigma  }(t')\right\rangle,
&\quad &
g_{\sigma \sigma'}F^{+}_{{z^2},3}({\bf k};t,t')=-i\left\langle T\
a^{(3)+}_{-{\bf k},\sigma}(t)c^{+}_{{\bf k},\sigma'}(t')\right\rangle, \nonumber
\end{eqnarray}
where $\hat g=i\hat \sigma^y$. Using the equations of motion for
operators $a^{(3)}_{{\bf k},\sigma}$ and $c_{{\bf k},\sigma}$ and
transforming to the frequency representation, we obtain the
following set of equations for the $z^2$ Green's functions:
\begin{equation}
\left( \omega -\varepsilon_{z^2}(\bf k)\right) G_{z^2}({\bf k},\omega)
-W_3({\bf k})G^{+}_{{z^2},3}({\bf k},\omega)+
\Delta _{z^2}F_{z^2}^{+}({\bf k},\omega)=1,
\end{equation}
\begin{equation}
\left( \omega -\varepsilon_{3}({\bf k})\right) G^{+}_{{z^2},3}({\bf k},\omega)
-W_3({\bf k})G_{z^2}({\bf k},\omega)+
\Delta _3({\bf k})F^{+}_{{z^2},3}({\bf k},\omega)=0,
\end{equation}
\begin{equation}
\left( \omega +\varepsilon_{z^2}(\bf k)\right) F_{z^2}^{+}({\bf k},\omega)
+W_3({\bf k})F^{+}_{{z^2},3}({\bf k},\omega)+
\Delta_{z^2}^{+}G_{z^2}({\bf k},\omega)=0,
\end{equation}
\begin{equation}
\left( \omega +\varepsilon_{3}({\bf k})\right) F^{+}_{{z^2},3}({\bf k},\omega)
+W_3({\bf k})F_{z^2}^{+}({\bf k},\omega)+
\Delta_3^{+}({\bf k})G^{+}_{{z^2},3}({\bf k},\omega)=0,
\end{equation}
    where
\begin{equation}
    \Delta_3({\bf k}) \equiv \Delta_{33}({\bf k})
=\displaystyle
\Delta_d \frac{(\varepsilon_p-\varepsilon_{3})^2}
{(\varepsilon_p-\varepsilon_{3})^2+(W_x)^2+(W_y)^2}\,.
\end{equation}

In the following, we will make use of quasiparticle density of
states of
 the
 $z^2$ band,
\begin{eqnarray} N(\omega)=-\frac{1}{\pi} \int Im\,G_{z^2}^R(\omega,{\bf k})\,
\frac{d^3 k}{(2 \pi)^3}\,.
\end{eqnarray} The retarded Green's
function is obtained by the solution of the system of equations
 (8-11) and has the form:
$$ G_{z^2}^R(\omega,{\bf k})=\frac{u^2_{-,{\bf k}}}{\omega -E_{-}({\bf k})+i 0}
+\frac{u^2_{+,{\bf k}}}{\omega -E_{+}({\bf k})+i
0}+\frac{v^2_{-,{\bf k}}}{\omega
 +E_{-}({\bf k})+i 0}+
\frac{v^2_{+,{\bf k}}}{\omega +E_{+}({\bf k})+i 0}\, ,$$ where the
coherence factors are defined as
$$ u^2_{\pm,{\bf k}}=\mp\frac{(E_-+\varepsilon_{z^2})(-E_-^2+\varepsilon_3^2+
\Delta_{3}^2)+W_3^2(E_--\varepsilon_3)}
{2E_-(E_{+}^2-E_{-}^2)}\,\ $$
$$ v^2_{\pm,{\bf k}}=\mp\frac{(E_--\varepsilon_{z^2})(-E_-^2+\varepsilon_3^2+
\Delta_{3}^2)+W_3^2(E_-+\varepsilon_3)}
{2E_-(E_{+}^2-E_{-}^2)}\,.
$$

The dispersion relations for two branches of the excitation
spectrum have the form

\begin{eqnarray}
 E_{\pm }^2({\bf k}) &=& \frac{\varepsilon _{z^2}^2({\bf k})+
\Delta _{z^2}^2+
\varepsilon_3^2({\bf k})+\Delta _{3}^2({\bf k})+2W_3^2({\bf k})}{2}
%\nonumber
\\
    &\pm & \frac{ \sqrt{(\varepsilon _{z^2}^2({\bf k})+\Delta _{z^2}^2-
\varepsilon_3^2({\bf k})-\Delta _{3}^2({\bf k}))^2+4W_3^2({\bf
k})\{(\Delta _{z^2}-\Delta_{3}({\bf k}))^2+ (\varepsilon
_{z^2}({\bf k})+\varepsilon_3({\bf k}))^2\}}}{2}. \nonumber
\end{eqnarray}

The set of equations (8-11) can be reduced to the system of two
equations for Green's functions $G_{z^2}$ and $F_{z^2}$ with the
effective order parameter defined by the formula
\begin{equation}\Delta_{z^2}({\bf k})=\frac{\Delta_3({\bf k}) W_3^2({\bf k})}
{\varepsilon_{3}^2({\bf k})+\Delta
_3^2({\bf k})}+\Delta_{z^2} .
\end{equation}
 The sign of
the order parameter $\Delta_3({\bf k})$(12) is determined by the
sign of the parameter $\Delta_d$(7), which can be positive or
negative depending on the type of interaction $U_d$ (repulsion or
attraction). The parameter $\Delta_d=0$ if $U_d=0$ or
$U_d>U^{crit}_d>0$, where $U^{crit}_d$ is a certain critical value
of repulsion on the $d_{x^2-y^2}$ orbital  of copper which
suppresses superconductivity in the system. As seen from the
formula (15), in the case of a nonzero interaction $U_d$ on the
$d_{x^2-y^2}$ orbital ($\Delta_d \ne 0$), the order parameter
depends on quasimomentum, and in the case of repulsion,
$\frac{\Delta_{3}({\bf k})} {\Delta_{z^2}} < 0 $ ($U_d>0$), the
parameter $\Delta_{z^2}({\bf k})$ changes its sign.

The approach considered explains the strong anisotropy of the
order parameter as a consequence of the symmetry properties of the
one-particle matrix element of the interband hybridization
$W_3({\bf k})$ \cite{ars}, which
 has nodes along the diagonals of the Brillouin
zone due to the difference between the types of symmetry of the
initial bands. In this case, the branch $E_{-}({\bf k})$ of the
excitation spectrum vanishes at points located on lines in the
${\bf k}$ space \cite{ding} at which the effective order parameter
$\Delta_{z^2}({\bf k})$ is equal to zero. It should be noted that
the initial interaction is regarded as isotropic; i.e., the effect
considered does not depend on the origin of the pairing mechanism.
In addition, the approach does not require a strong anisotropy of
the spectrum and, hence, can be used for various types of
high-temperature superconductors. The model proposed in \cite{com}
also leads to formulas of the type (14), (15), but it is based on
the exotic condition of the interaction sign reversal in various
regions of the Fermi surface.

The diagonalization of the one-particle part of Hamiltonian (1)
leads to the problem with anisotropic attraction in energy bands.
The effective order parameters in the bands may have nodes, but
have no pure  $d$ or $(s + d)$ symmetry since their anisotropy is
determined by the band representations of the space symmetry group
of the lattice \cite{ars,vol,sok}.

 The formulated model of the order parameter
anisotropy makes it possible to obtain the $\displaystyle
\frac{dI}{dV}$ characteristics of $SIS$ junctions for
superconductors of the BSCCO type. Taking into account the crystal
structure of these compounds, we will assume that tunneling along
the $c$ axis occurs mainly through the $d_{z^2}$ orbitals of
copper in the $CuO_2$ plane (and through the apical oxygen which
is not included explicitly in the model). The matrix element of
tunneling is assumed to be independent of momentum ($T_{\bf
kp}=T=const$) due to  the random formation of bonds between the
$d_{z^2}$ orbitals on both sides of the break junction. Thus,
tunneling between two superconductors ($CuO_2$ layers) is treated
as occuring through a number of point junctions.

The expression for the dependence of the quasiparticle tunnel
current on the voltage applied to the junction in this case
assumes the standard form:
\begin{eqnarray} I(V)=4e|T|^2
\int\limits_{-\infty}^{\infty}[n(\omega)-n(\omega-eV)]N(\omega)
N(\omega-eV)d\omega\,, \end{eqnarray} where
$n(\omega)=[e^{\textstyle\frac{\omega}{T}}+1]^{-1}$
 is the Fermi
distribution function and $N(\omega)$ is defined by formula (13).
All calculations were performed for temperature $T = 0$. An
important aspect of this approach is that $N(\omega)$ is not the
average density of states in the conduction band (as in the
single-band BCS model), but is a partial density on the $d_{z^2}$
orbital.

 In order to find the
density of states $N(\omega)$ and the tunnel conductance
$\displaystyle \frac{dI}{dV}$ in accordance with the chosen
approximation, we used the values of parameters from the
tight-binding model, which correspond to the calculations of the
band structure of HTSC \cite{das}. If all parameters are expressed
in terms of $\Delta_{z^2}$ and the position of energy levels is
measured from the chemical potential, we have $t_{d-p}=75$,
$t_{z^2-p}=15$, $\varepsilon_d=-75$ and $\varepsilon_p=-50$. In
all calculations, we assumed the existence of a finite relaxation
constant $\gamma=0.05$. The parameter determining the width of the
initial $\varepsilon_{z^2}(\bf k)$
 band is  $t_{z^2}=1.5$.

For the chosen values of the parameters, we analyzed the behavior
of the characteristics $N(\omega)$ and $\displaystyle
\frac{dI}{dV}$ depending on the interaction value between
electrons on the $d_{x^2-y^2}$ orbital of copper and on the
position of the center $\varepsilon_{z^2}$ of the initial $z^2$
band relative to the chemical potential (Figs. 2-4). The obtained
dependencies were compared with these of single-band model with
the $s$- and $d$-types of symmetry of the order parameter
($\Delta=const$ and $\Delta \sim(cos(k_x)-cos(k_y))$,
respectively).

Figure 2A shows the $N(\omega)$ curves calculated for the normal
state ($\Delta_{z^2}=0$). These curves display split peaks
associated with the Van Hove singularities of the initial
$\varepsilon_{z^2}({\bf k})$ band for various positions of its
center relative to  the chemical potential, as well as the
following peaks formed by singularities of the $\varepsilon_3({\bf
k})$ band. Figure 2b shows the same curves in the case when the
$\varepsilon_{z^2}({\bf k})$ spectrum is  replaced by a
dispersionless level ($t_{z^2}=0$).

\begin{figure}
%\leavevmode \centering{
 \centering{\includegraphics{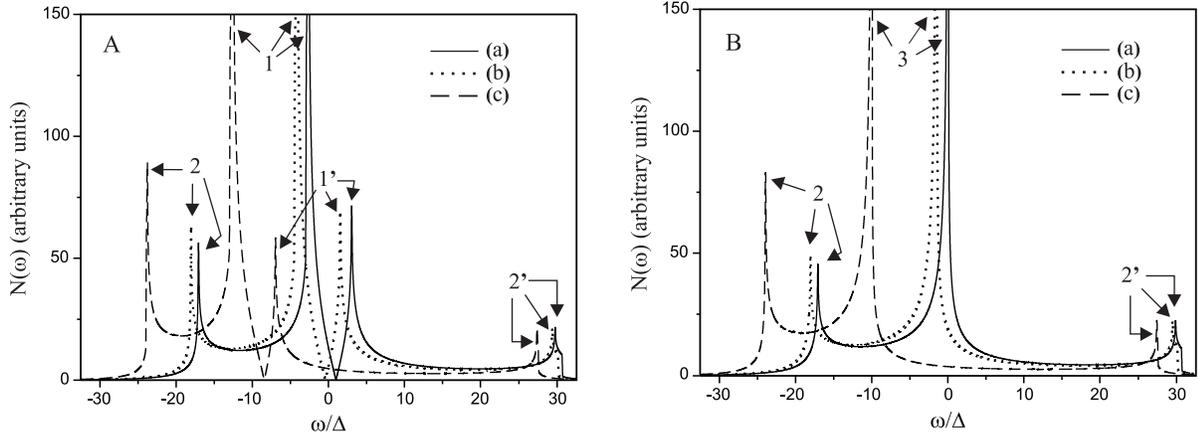}} \caption{
Normal density of states in the $z^2$ band: A) $t_{z^2}=1.5$, B)
$t_{z^2}=0$ at $\varepsilon_{z^2}=0$ (curves $a$), $-1.5$ (curves
$b$) and $-10$ (curves $c$). Peaks 1 and 1' are due to the Van
Hove singularities of the initial $z^2$ band, peaks 2 and 2' are
formed by the Van Hove singularities in the $\varepsilon_3({\bf
k})$ band, and peaks 3 are associated with the $\varepsilon_{z^2}$
level. }
\end{figure}

\begin{figure}[h]
%\leavevmode \centering{
\centering{\includegraphics{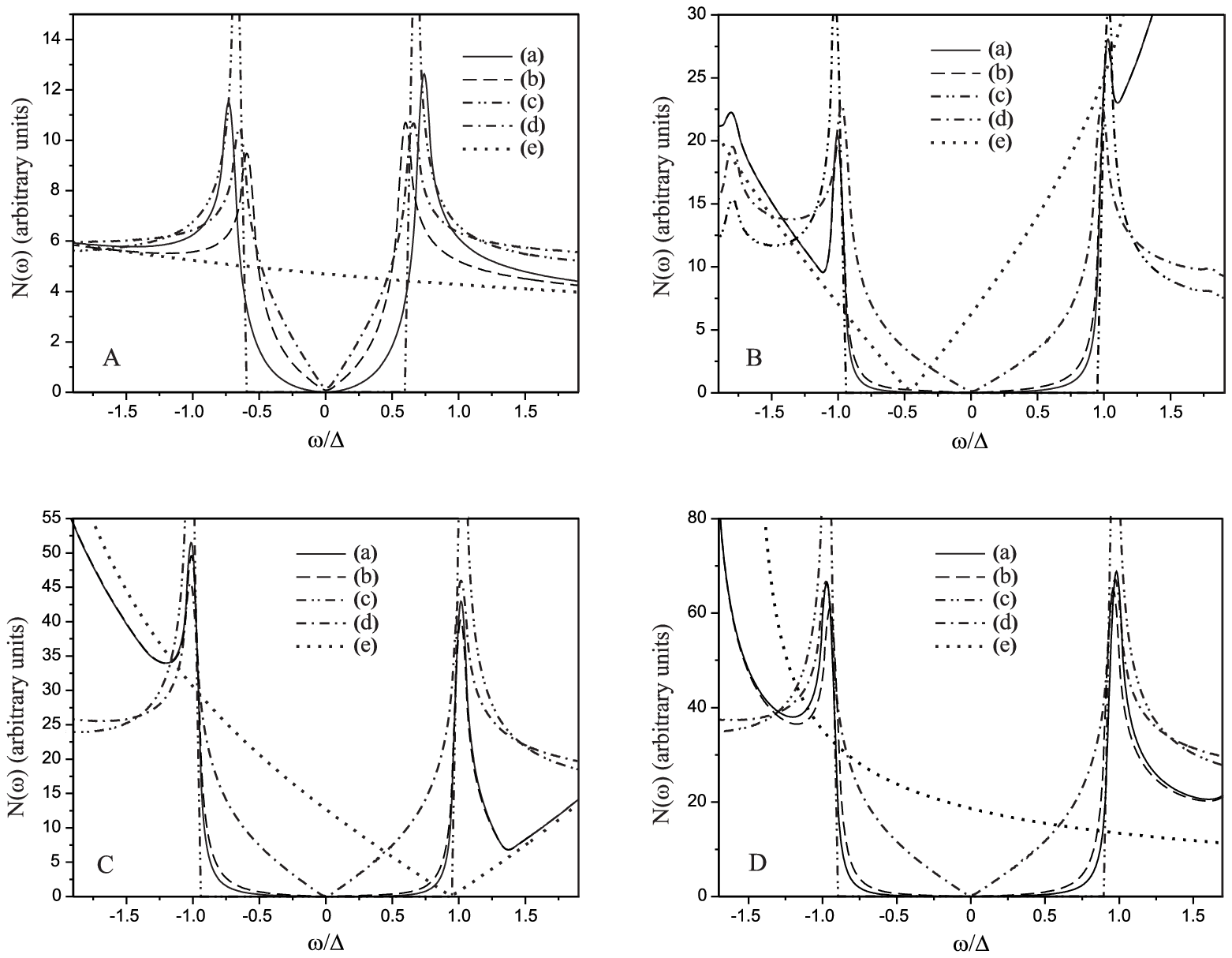}} \caption{
Density of states in the normal and superconducting states in the
$z^2$ band:
 A) $\varepsilon_{z^2}=-10$, $t_{z^2}=1.5$,
B) $\varepsilon_{z^2}=-1.5$, $t_{z^2}=1.5$, C)
$\varepsilon_{z^2}=0$, $t_{z^2}=1.5$, D) $\varepsilon_{z^2}=-1.5$,
$t_{z^2}=0$.
 Curves $a$ are plotted in the absence of interaction on
the $d_{x^2-y^2}$ orbital ($U_d=0$); curves $b$ take into account
the repulsion on the $d_{x^2-y^2}$ orbital ($U_d>0$); curves $c$
plotted in a model with the $s$ symmetry of the order parameter,
curves $d$ are plotted in a model with the $d$ symmetry of the
order parameter;  curve $e$ illustrates the normal density of
states.}
\end{figure}

Fig.3.A presents the $N(\omega)$ dependencies in the
superconducting state for the case when the $z^2$ band is far from
the chemical potential ($\varepsilon_{z^2} = - 10$). It can be
seen from the figure that if there is electron-electron repulsion
on the $d_{x^2-y^2}$ orbital $\Delta_d=- 0.5$), the density of
quasiparticles is similar to that calculated in a model with the
$d$ symmetry of the order parameter (in particular, it is a linear
function of $\omega$ in the vicinity of $\omega=0$). If there is
no interaction on the $d_{x^2-y^2}$ orbital, the distance between
the superconducting peaks and their height increase. According to
analytical estimations, $N(\omega)\sim \omega^{3/2}$ for very
small frequencies. Similar results can also be obtained in the
single-band model with a nontrivial spectrum and an effective
order parameter of certain symmetry. As noted above, taking into
account the above approximations, the diagonalization of the
single-particle part of
 Hamiltonian (1) leads to the single-band model with anisotropic pairing. In this
case, the symmetry of the corresponding order parameter can be
approximated by an $(s+d)$-type symmetry.

If the center of the initial $z^2$ band is close to the chemical
potential (see Figs. 3B and 3C), the difference between the $a$
and $b$ curves (with and without taking into account of repulsion)
vanishes due to the dominant role of the Van Hove singularities on
energy scales of the order of ${\Delta_{z^2}}$. In both cases, the
behavior of the density of quasiparticles becomes of the $s$ type.
The calculated density of states demonstrates the asymmetry of
peaks, which is associated with the position of Van Hove
singularities in the normal density of states (see Fig. 2). The
Van Hove singularity nearest to the chemical potential increases
the height of the corresponding peak in the density of states in
the superconducting state. A comparison of Figs. 3B and 3C shows
that, for a certain intermediate value of $\varepsilon_{z^2}$,
there is the mirror switching of the asymmetry of peaks associated
with the model spectrum structure.

The curves presented in Fig. 3A show that even without the
repulsion on the $d_{x^2-y^2}$- orbital if the $\varepsilon_{z^2}$
level is far from the chemical potential
($\varepsilon_{z^2}=-10$), the distance between the
superconducting peaks is smaller than $2\Delta_{z^2}$. As the
$\varepsilon_{z^2}$ level approaches the chemical potential, the
distance between the superconducting peaks increases, and the
position of the peaks on the $N(\omega)$ curve for
$\varepsilon_{z^2}=0$ corresponds to the parameter $\Delta_{z^2}$.
In this case, the height of the peaks increases significantly. The
distance between the peaks changes as a result of the displacement
of the $\varepsilon_{z^2}$ level relative to the chemical
potential for the fixed values of the remaining parameters.
However, a more detailed analysis of this phenomenon requires the
solution of the self-consistent equations for order parameters as
functions of the parameters of the model as well as the
determination of the interaction constants $U_{z^2}$ and $U_d$
taking into account the specific mechanism of pairing, which is
beyond the scope of the present paper. The dependence of the
distance between the superconducting peaks on the position of the
chemical potential in the model with the $(s + d)$-type of order
parameter symmetry  was also found theoretically in \cite{cof}.

A comparison of the $N(\omega)$ dependencies found in the model
with single-band models with the $s$- and $d$-types of the order
parameter symmetry was carried out taking into account the
location of the Van Hove singularities associated with the initial
$\varepsilon_{z^2}({\bf k})$ band. This is pointed out by similar
asymptotic behavior of densities of states calculated for both the
normal and the superconducting state in $s$-, $d$- and considered
models (see Fig. 3C) on the scales of energy higher than
$\Delta_{z^2}$.

A comparison of the results presented in Figs. 3B and 3D shows
that, as the width of the initial ${z^2}$ band determined by the
parameter $t_{z^2}$ decreases ($t_{z^2}=1.5$ in Fig.3B and
$t_{z^2}=0$ in Fig.3D), the heights of the superconducting peaks
increases, since the singularity in the normal density of states
in this case is displaced towards the chemical potential (see Fig.
2).

\begin{figure}[h]
%\leavevmode \centering{
\centering{\includegraphics{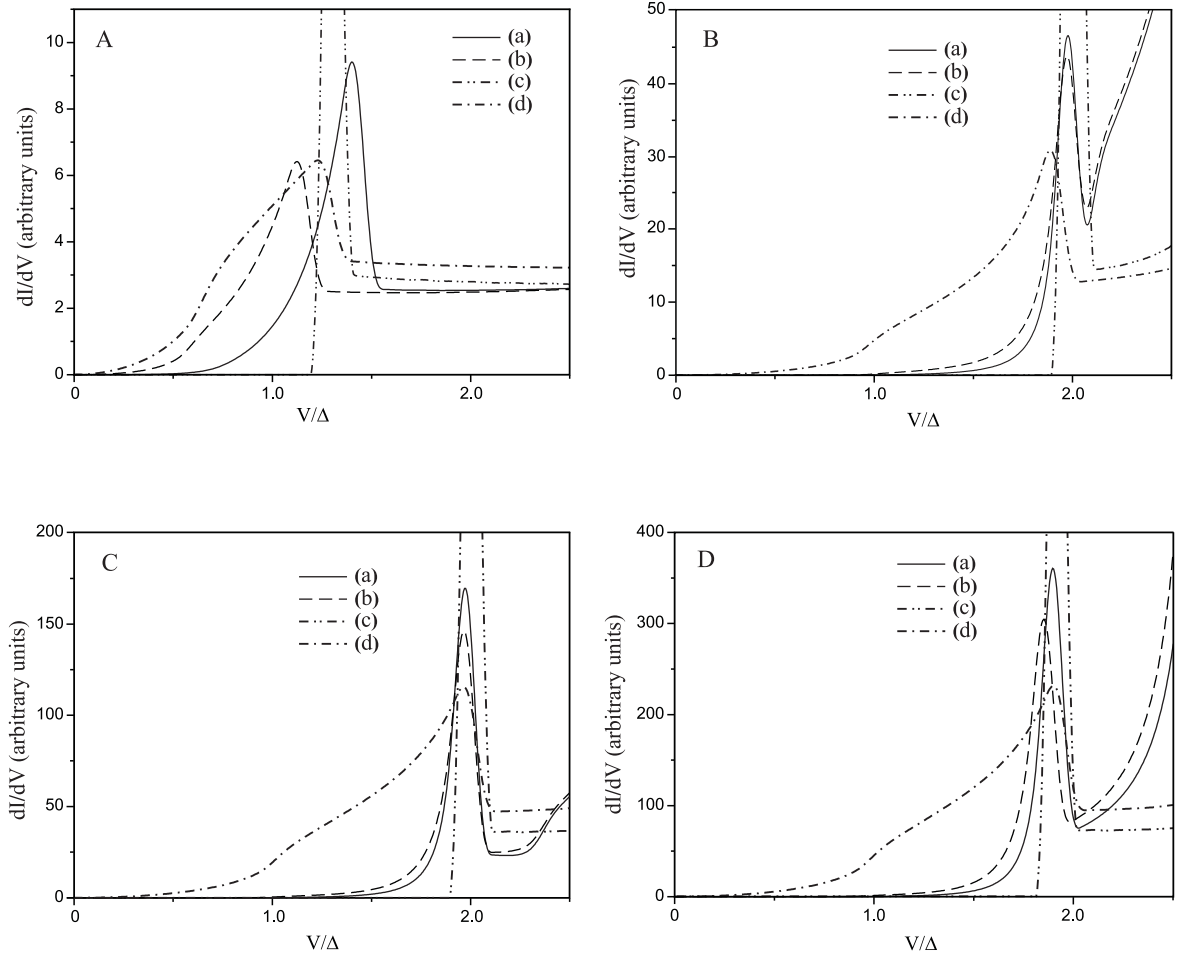}} \caption{
$\displaystyle \frac{dI}{dV}$ characteristics for A)
$\varepsilon_{z^2}=-10$, $t_{z^2}=1.5$, B)
$\varepsilon_{z^2}=-1.5$, $t_{z^2}=1.5$, C) $\varepsilon_{z^2}=0$,
$t_{z^2}=1.5$, D) $\varepsilon_{z^2}=-1.5$, $t_{z^2}=0$.
 Curves $a$
are plotted in the absence of interaction on the $d_{x^2-y^2}$
(($U_d=0$); curves $b$ take into account the repulsion on the
$d_{x^2-y^2}$ orbital ($U_d>0$); curves $c$ plotted in a model
with the $s$ symmetry of the order parameter, curves $d$ are
plotted in a model with the $d$ symmetry of the order parameter.}
\end{figure}

Figure 4 shows the $\displaystyle\frac{dI}{dV}$ characteristics of
$SIS$ junction calculated by formula (16) using the determined
$N(\omega)$ dependencies for three positions of the
$\varepsilon_{z^2}$ level relative to the chemical potential
$\varepsilon_{z^2}=-10$, $\varepsilon_{z^2}=-1.5$,
$\varepsilon_{z^2}=0$, which can correspond to different HTSC
types. The common feature of these characteristics is that in the
case of repulsion the superconducting peak appears at lower
voltages and is lowered, but the curve at low voltages lies higher
than in the absence of interaction on the $d_{x^2-y^2}$ orbital of
copper. If the center of the initial $z^2$ band is far from the
chemical potential ($\varepsilon_{z^2}=-10$), the
$\displaystyle\frac{dI}{dV}$ characteristics are similar to those
calculated in a model with the $d$ symmetry of the order parameter
(Fig. 4a). As the $\varepsilon_{z^2}$ level approaches the
chemical potential, the curves tend to the dependencies obtained
in the model with the $s$ symmetry of the order parameter (Figs.
4b and 4c). For low voltages $V<\Delta_{z^2}$, the $IV$
characteristic of an $SIS$ junction is strongly suppressed. In
analogy with the $N(\omega)$ dependencies, the
$\displaystyle\frac{dI}{dV}$ characteristics exhibit extremely
narrow superconducting peaks. This corresponds to a number of
experimentally observed results for BSCCO-type compounds, e.g.,
optimally doped \cite{pon}, overdoped \cite{suz}, and underdoped
\cite{dew} $Bi_2Sr_2CaCu_2O_{8+\delta}$.

Thus, the electron density of states and current-voltage
characteristics of $SIS$ junctions are calculated in the framework
of the multiband HTSC model with an anisotropic effective order
parameter. We took into account the hybridization between the
$d_x$ and $d_y$ orbitals of oxygen and the $d_{x^2-y^2}$ orbitals
of copper in the $CuO_2$ plane and assumed the presence of
electron-electron attraction on the $d_{z^2}$ orbital of copper.
The performed calculations demonstrate the possibility of
explanation the results of experimental measurements of the
electron density of states and the tunnel properties of $SIN$ and
$SIS$ junctions (e.g., $s$-type characteristics of BSCCO-type
compounds) on the basis of the model taking into account the
structure of the electron spectrum of HTSC. It should be stressed
that anisotropy of the order parameter and of the excitation
spectrum is explained only by the symmetry of the crystal lattice
and atomic orbitals and does not depend on the nature of pairing.

The main result of this work is that for a strongly anisotropic
order parameter $\Delta(\bf k)$ (15) (e.g. of $d$-type) and in the
presence of nodes in the excitation spectrum of a superconductor,
the simple inclusion of the real band structure of the HTSC in a
wide range of the model parameter values results in the
experimentally observed \cite{pon,suz,dew}
 $s$-type behavior of the current-voltage characteristics
$SIN$ and $SIS$ junctions. A comparison of the curves calculated
for various values of the parameters of the model and in the
simplest cases of models with the $s$ and $d$ symmetries of order
parameters indicates a strong dependence of both the density of
states and the $\displaystyle\frac{dI}{dV}$ characteristics on the
electron band structure of high-temperature superconductors. These
characteristics also depend on the crystal lattice symmetry and on
the presence of an additional interaction between electrons, which
was introduced in the model on the $d_{x^2-y^2}$ orbital of
copper. Above analysis points to the necessity of considering of
the band structure of HTSC for interpretation of the results of
tunnel experiments .

% \begin{acknowledgments}
 The authors are
grateful to B.A. Volkov, E.G. Maksimov, and Ya.G. Ponomarev for
numerous discussions in the course of the research. This work was
supported by the Russian Foundation for Basic Research (N
02-02-16925, N 00-15-15998) and Landau Scholarship Grant.
%\begin{acknowledgments}

%\begin{references}

\end{document}